\documentclass[twocolumn]{aastex7}
\usepackage{subcaption} 
\usepackage{caption}    

\usepackage{xspace}
\usepackage{amsmath}

\newcommand{\teff}{\ensuremath{T_\mathrm{eff}}}

\newcommand{\fsed}{\ensuremath{f_\mathrm{sed}}}

\newcommand{\fpatch}{\ensuremath{F_\mathrm{patch}}}

\newcommand{\fbase}{\ensuremath{F_\mathrm{base}}}

\newcommand{\ftot}{\ensuremath{F_\mathrm{total}}}

\begin{document}

\title{Gray Spectral Variability in Three Brown Dwarfs Observed by HST/WFC3 Time-Series Observations}

\correspondingauthor{Madalyn F. Chapleski, Yifan Zhou}
\email{ppy2ej@virginia.edu, yzhou@virginia.edu}

\author[0009-0005-7812-813X]{Madalyn F. Chapleski}
\affiliation{Department of Astronomy, University of Virginia, 530 McCormick Rd., Charlottesville, VA 22904, USA}
\email{ppy2ej@virginia.edu}

\author[0000-0003-2969-6040]{Yifan Zhou}
\affiliation{Department of Astronomy, University of Virginia, 530 McCormick Rd., Charlottesville, VA 22904, USA}
\email{yzhou@virginia.edu}



\begin{abstract}

The L/T transition is a critical evolutionary stage for brown dwarfs and self-luminous giant planets. L/T transition brown dwarfs are more likely to be spectroscopically variable, and their high-amplitude variability probes distributions in their clouds and chemical makeup. This paper presents Hubble Space Telescope Wide Field Camera 3 spectral time series data for three variable L/T transition brown dwarfs and compares the findings to the highly variable benchmark object 2MASS J2139. All four targets reveal significant brightness variability between 1.1 to 1.65 micron but show a difference in wavelength dependence of the variability amplitude. Three of our targets do not show significant decrease in variability amplitude in the 1.4 $\mu$m water absorption band commonly found in previous studies of L/T transition brown dwarfs. Additionally, at least two brown dwarfs have irregular-shaped, non-sinusoidal light curves. We create heterogeneous atmospheric models by linearly combining SONORA Diamondback model spectra, comparing them with the observations, and identifying the optimal effective temperature, cloud opacity, and cloud coverage for each object. 
Comparisons between the observed and model color-magnitude variations that trace both spectral windows and molecular features reveal that the early- T dwarfs likely possess heterogeneous clouds. The three T dwarfs show different trends in the same color-magnitude space which suggests secondary mechanisms driving their spectral variability. This work broadens the sample of L/T transition brown dwarfs that have detailed spectral time series analysis and offers new insights that can guide future atmospheric modeling efforts for both brown dwarfs and exoplanets. 

\end{abstract}



\section{Introduction} \label{sec:intro}

Shaped by a variety of dynamical processes, the atmospheres of brown dwarfs are heterogeneous and variable \citep{Showman2013,Zhang2020,Tan2021a,Tan2021b}. These heterogeneous structures are manifested in the spatial distributions of thermal profiles, condensate cloud coverage, and possibly distributions of carbon-bearing molecules \citep{Marley2010,Robinson2014,Tremblin2020}. These variations appear as periodic rotational modulations and irregular light curve evolution \citep{Metchev2015,Apai2017,Biller2024,McCarthy2025}. Brown dwarfs resemble directly imaged planets in their temperature and compositions \citep{Faherty2016} and young and low-mass brown dwarfs also share similar surface gravity to directly imaged exoplanets \citep{Gagne2017,Zhang2025}. Unlike planets, the spectroscopic observations of brown dwarfs are free from the contamination of bright host stars. Consequently, time series observations of brown dwarfs are effective probes of dynamic atmospheric processes in extrasolar substellar atmospheres.


The L/T transition region has been a focus of brown dwarf variability studies \citep[e.g.,][]{Kirkpatrick2005,Radigan2014,Metchev2015,Vos2019}. This region exhibits a dramatic change in near-infrared colors in the color-magnitude diagram (CMD) from red L-dwarfs to blue T-dwarfs \citep{Kirkpatrick2005}. At these spectral types, iron and silicate condensates form near the bottom of the near-infrared photosphere \citep[e.g.,][]{Ackerman2001}. Cloud models suggest that as condensation front sinks deeper in cooler atmospheres, the horizontal variations become more likely because of increased turbulence motions  \citep{Ackerman2001}. This heterogeneous cloud structure could explain both the observed color change and frequent occurrence of high-amplitude rotational modulations in the L/T transition types \citep[e.g.,][]{Marley2010}. However, \citet{Lew2020} analyzed a sample of brown dwarfs and found that the $J-H$ color variations are mostly gray. They concluded that the dramatic color-magnitude changes between L and T spectral types (several magnitudes) have a different physical origin than the subtle variations observed in individual objects (typically less than a few percent).

The L/T transition also marks the point where methane becomes a significant opacity source \citep{Kirkpatrick1999}. Atmospheric chemical modeling suggests that the CO-CH$_4$ reaction, coupled with vertical mixing and convection, can drive fingering convection \citep{Tremblin2016}. This process creates local variations in methane abundance and temperature structure. Combined with the patchy cloud formation discussed above, these chemical and dynamical processes can produce spectral variability \citep{Morley2014,Tremblin2020}. The interplay between clouds and chemistry may therefore manifest as complex wavelength-dependent variability in both amplitude and phase.

Over the past decade, Hubble Space Telescope Wide Field Camera 3 (HST/WFC3) was the primary space-based instrument for characterizing brown dwarf spectral variability \citep[e.g.,][]{Buenzli2012,Apai2013,Biller2018,Lew2016,Lew2020,Zhou2020}. Recently, James Webb Space Telescope (JWST) time series observations have provided exceptional results with superior signal-to-noise ratios and broader wavelength coverage \citep{Biller2024,McCarthy2025}. However, HST/WFC3 data remain invaluable due to their extensive volume, which enables statistical exploration of the diversity in brown dwarf variability behavior. We have systematically explored the HST archive to identify unpublished time series spectroscopic observations of brown dwarfs and present our comprehensive analysis in this paper.

We present a new analysis of three L/T transition brown dwarfs and compare their variability characteristics to the well-studied, highly variable benchmark object 2MASS J21392676+0220226 (2MASS J2139). We use data originally published on 2MASS J2139 in \cite{Apai2013} which observes a variability amplitude of $\sim$27\%.
\cite{Vos2017} detects 2MASS J2139's inclination to be 90$^\circ$, making it the ideal viewing geometry to detect equatorial variability. In their analysis of 2MASS J2139, \cite{Apai2013} showed how its spectral variations can only be described by a linear combination of two types of spectra, hinting that it has a homogeneous atmosphere except for a single atmospheric feature. 2MASS J2139's large variability amplitude makes it an ideal candidate to compare to less variable objects, as its pronounced spectral changes provide a clear benchmark for understanding the atmospheric processes driving variability in brown dwarfs. 

\cite{Burgasser2008} classified 2MASS J11263991-5003550 (2MASS J1126) as a blue L-dwarf due to its normal optical spectra but characteristically blue near-infrared colors. The authors attribute the blue near-infrared spectrum to the presence of thin, uniform condensate clouds in the atmosphere, along with potentially high surface gravity and the presence of subsolar metallicities \citep{Burgasser2008}. \cite{Marley2010} conducted a more recent analysis of 2MASS J1126's atmosphere and proposed that its atmosphere is composed of thicker patchy clouds as opposed to the thin homogeneous cloud cover model suggested by \cite{Burgasser2008}. \cite{Marley2010} discusses the difficulty in distinguishing globally homogeneous atmospheres from partly cloudy ones due to their similar spectra. Thus, a more in depth analysis of 2MASS J1126's atmosphere is necessary to determine the nature of its cloud structure. However, with a viewing inclination of 35$^\circ$, 2MASS J1126 is not ideally oriented for accurately measuring face-on variability \citep{Vos2017}.

Lastly, 2MASS J07584037+3247245 (2MASS J0758) and 2MASS J16291840+0335371 (2MASS J1629) are early T-dwarfs with significant variability amplitudes of 4.8\% and 4.3\% detected in \cite{Radigan2014}. A past survey performed by \cite{BardalezGagliuffi2015} identified 2MASS J0758 as a potential binary based on visual inspection. Additionally, a previous study with a ground based telescope found 2MASS J1629 exhibits $\sim$10\% variability that potentially evolves from night to night \citep{Heinze2015}. Similar to 2MASS J2139, 2MASS J1629 has a viewing inclination of 82$^\circ$, making it ideal for detecting rotationally modulated variability features \citep{Vos2017}.

In this paper, we present previously unanalyzed HST/WFC3 light curves on 2MASS J1126, J1629, and J0758 and compare our findings to the highly variable benchmark object 2MASS J2139 \citep{Apai2013}. We model each object's variability using a linear combination of atmospheric models produced by SONORA Diamondback \citep{Morley2024} and use the same model grid to analyze the trajectories of the objects in color-magnitude space. By comparing these trajectories with model predictions, we infer the physical mechanisms driving their variability. The structure of the paper is as follows: we present the observations and the methods of data reduction in Section \ref{observations}, describe the variability and display the light curves in Section \ref{variability}, present the results of atmospheric modeling in Section \ref{modeling}, analyze the trajectory of each object relative to the model grid in color-magnitude space and infer the mechanisms behind their variability in Section \ref{cmd}, and summarize our results in Section \ref{summary}.

\section{Observations and Data Reduction} \label{observations}

\begin{deluxetable*}{ccccc}
\tablecaption{Data collection information for each target.\label{tab:obsdata}}
\tablewidth{0pt}
\tablehead{
\colhead{Brown Dwarf} & \colhead{Date Observed} & \colhead{Time Range} & \colhead{Maximum Cadence} & 
\colhead{Average S/N} \\
\colhead{} & \colhead{} & \colhead{hr} & \colhead{s} & 
\colhead{}}
\startdata
2MASS J1126 & 2013-11-1 to 11-2 & 4.175 & 141.8 & 162.2 \\
2MASS J1629 & 2015-6-6 & 5.420 & 247.0 & 81.38 \\
2MASS J0758 & 2014-4-12 & 7.107 & 232.6 & 111.8 \\
2MASS J2139 & 2010-10-21 & 8.631 & 478.0 & 44.35
\enddata
\end{deluxetable*}


\begin{deluxetable*}{cccccccccc}
\tablecaption{Target properties and time-average photometric measurements. \label{tab:backgroundinfo}}
\tablewidth{0pt}
\tablehead{
\colhead{Brown Dwarf} & \colhead{Spectral Type} & \colhead{$\teff$} & \colhead{Inclination} & \colhead{F127M} & \colhead{F139M} & \colhead{F153M} & \colhead{F127M-F153M} & \colhead{F127M-F139M} \\
\colhead{} & \colhead{} & \colhead{K} & \colhead{$^\circ$} & 
\colhead{mag} & \colhead{mag} & \colhead{mag} & \colhead{mag} & \colhead{mag}}
\startdata
2MASS J1126 & L6.5 & 1679 & 35 & 13.71 & 14.62 & 13.66 & $0.051$ & $-0.912$\\
2MASS J1629 & T2 & 1085 & 82 & 14.88 & 16.83 & 14.91 & $-0.032$ & $-1.949$\\
2MASS J0758 & T2 & 1107 & \nodata & 14.55 & 16.41 & 14.50 & $0.042$ & $-1.865$\\
2MASS J2139 & T1.5 & 1060 & 90 & 14.64 & 16.32 & 14.76 & $-0.122$ & $-1.682$ \\
\enddata
\tablecomments{Spectral type, \teff, and spin axis inclination are from \citet[][]{Radigan2014}, \citet{Burgasser2008}, \citet{Vos2017}, and \citet{Best2024-oc}.}
\end{deluxetable*}

HST/WFC3 G141 grism spectral timeseries of brown dwarfs 2MASS J0758, J2139, J1629, and J1126 were collected on 2010 October 21, 2013 November 1-2, 2014 April 12, and 2015 June 6, respectively. The observations of 2MASS J2139, previously analyzed by \cite{Apai2013}, are presented here solely as a reference due to its large variability amplitude. The low-resolution ($R\sim130$) spectra span from 0.968 to 1.805 $\mu$m, encompassing the prominent water absorption band centered at 1.4 $\mu$m. Our analysis only includes the 1.1 to 1.65 $\mu$m range, where the spectral throughput is above 90\% of the maximum and the signal-to-noise ratios (S/N) of the extracted spectra are high.

Observations of 2MASS J1126, 2MASS J1629, 2MASS J0758, and 2MASS J2139 consists of three, four, five, and six consecutive HST orbits, resulting in time baselines of 4.2 to 8.6 hours. In each orbit, the observation started with several short direct-imaging frames for wavelength calibration purposes, which were followed by consecutive spectroscopic exposures. These observations deliver 107, 80, 111, and 66 frames for 2MASS J1126, J1629, J0758, and J2139, respectively. The exposure times and cadence are set individually based on the brightness of the targets. Table~\ref{tab:obsdata} provides the details of the observations. Table \ref{tab:backgroundinfo} gives relevant background information on each target  and provides the observed magnitudes and colors using the F127M, F139M, and F153M filters.

Data reduction starts with the \texttt{flt} files downloaded from the MAST archive. Basic calibration steps, including dark and bias correction, flat field, and up-to-ramp fit, have been conducted by the \texttt{calwf3} pipeline. The time-resolved spectra are extracted by an \texttt{aXe}-based pipeline, which has been widely adopted in time-resolved HST/WFC3 spectroscopic data analysis \citep[e.g.,][]{Apai2013,Lew2020}. A $r=4$\,pixels radius aperture is adopted for spectral extraction. The pipeline provides time-resolved spectra and their associated uncertainties in both count rate (e$^{-}$\,s$^{-1}$) and $f_\lambda$ (erg\,s$^{-1}$\,cm$^{-2}$\,\AA$^{-1}$) units. Uncertainties were calculated by propagating pixel-level noise (photon noise, read noise, and calibration uncertainties) to the extracted spectrum. The ramp-effect models were created for individual time series data using physically based model RECTE \citep{Zhou2017}. After dividing the RECTE model profiles, the ramp effect systematics were removed from the data, including the first orbits.
The extracted spectra have average S/N values ranging from 50 to 150 per wavelength unit (Table~\ref{tab:obsdata}). 


\section{Spectral Variability Characterization} \label{variability}
\subsection{Spectral Time Series}
\begin{figure*}[ht!]
        \centering
        \includegraphics[width=\textwidth]{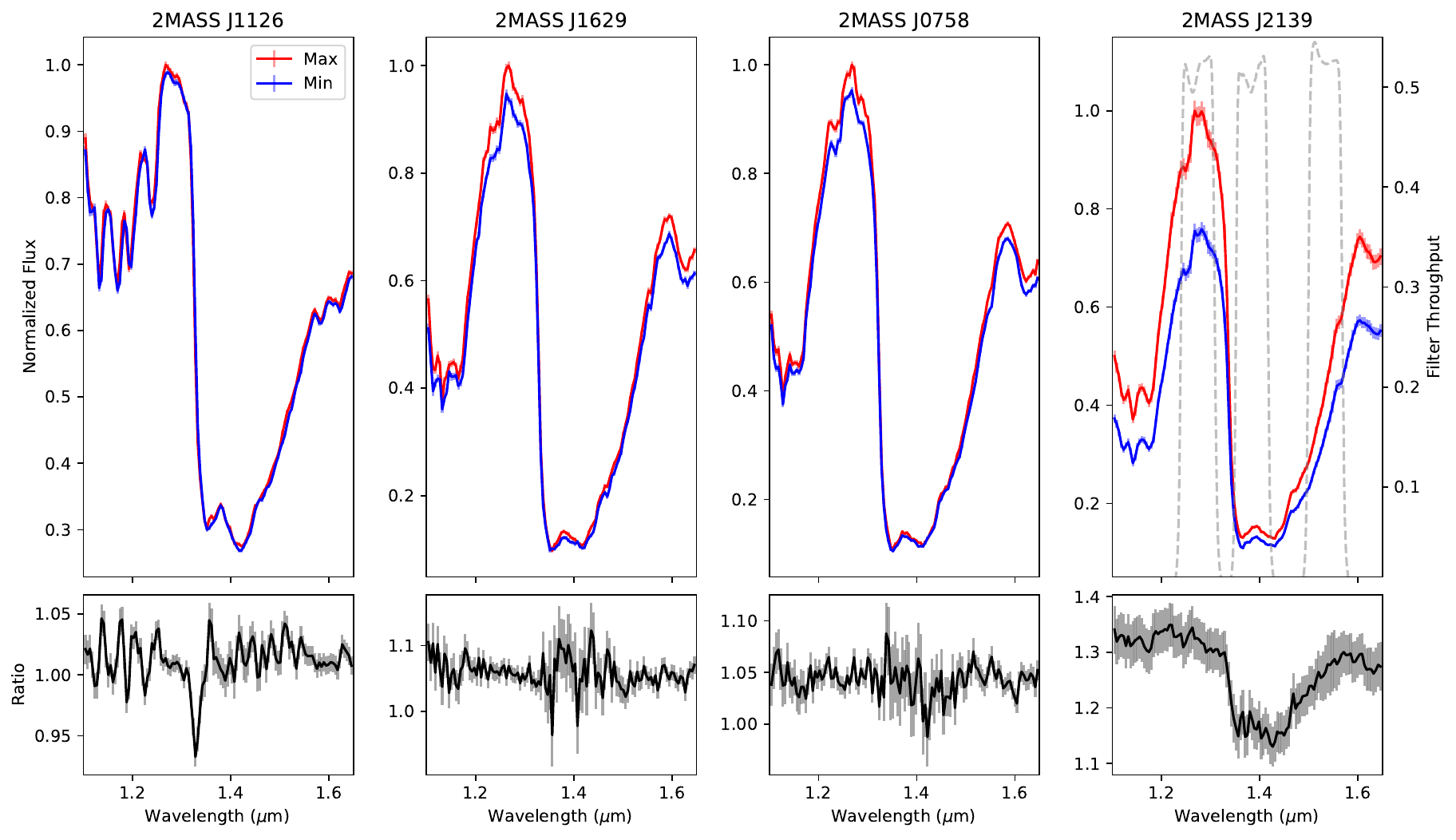} 
        \caption{Spectral variability detected in all three new targets (first three columns) and verified in the comparison source 2MASS J2139 (the last column). \textit{Top panel:} Normalized minimum (blue) and maximum (red) spectra with uncertainties. They are normalized to the highest flux value of each target's maximum spectrum. Transmission curves for the F127M, F139M, and F153M filters used for synthetic photometry in later sections are shown for 2MASS J2139. \textit{Bottom panel:} Maximum-to-minimum flux ratios with uncertainties. Target names are shown in the top panel title. The dip in flux ratio at $\sim$1.35 $\mu$m for 2MASS J1126 is a systematic feature due to the pointing drift of the telescope and should not be interpreted as a normal signal.}
        \label{fig:minmaxratio}
    \end{figure*}

The three newly analyzed brown dwarfs exhibit significant spectral variability, albeit with smaller amplitudes than 2MASS J2139 (Figure~\ref{fig:minmaxratio}). Figure~\ref{fig:minmaxratio} shows the maximum and minimum spectra and their ratios, with extrema selected by total integrated flux. A key difference from 2MASS J2139 is that 2MASS J1126, J1629, and J0758 show largely wavelength-independent (gray) variability across the $1.10$ to $1.65\,\micron$ range.

We define the peak-to-peak variability amplitude as $(F_\mathrm{max}/F_\mathrm{min}-1)\cdot100\%$, where $F_{\mathrm{max}}$ and $F_\mathrm{min}$ are the flux values averaged over our 1.10--1.65\,\micron{} analysis range for the maximum and minimum spectra. This yields variability amplitudes of 1.12 $\pm$ 0.09\%, 4.47 $\pm$ 0.11\%, and 5.80 $\pm$ 0.15\% for 2MASS J1126, 2MASS J0758, and 2MASS J1629, respectively, in good agreement with previous $J$ and $K$ band photometric measurements \citep{Radigan2014}.

For 2MASS J1126, telescope pointing drift caused a wavelength shift. We corrected this using cross-correlation and shifting the spectral image before extraction. Despite careful calibration, residual systematics remain near the water absorption feature (Figure \ref{fig:minmaxratio}), but these do not affect our results since our analysis excludes this spectral region.

\subsection{Light Curve Analysis}

\begin{figure*}[t]
    \centering
    \includegraphics[width=\textwidth]{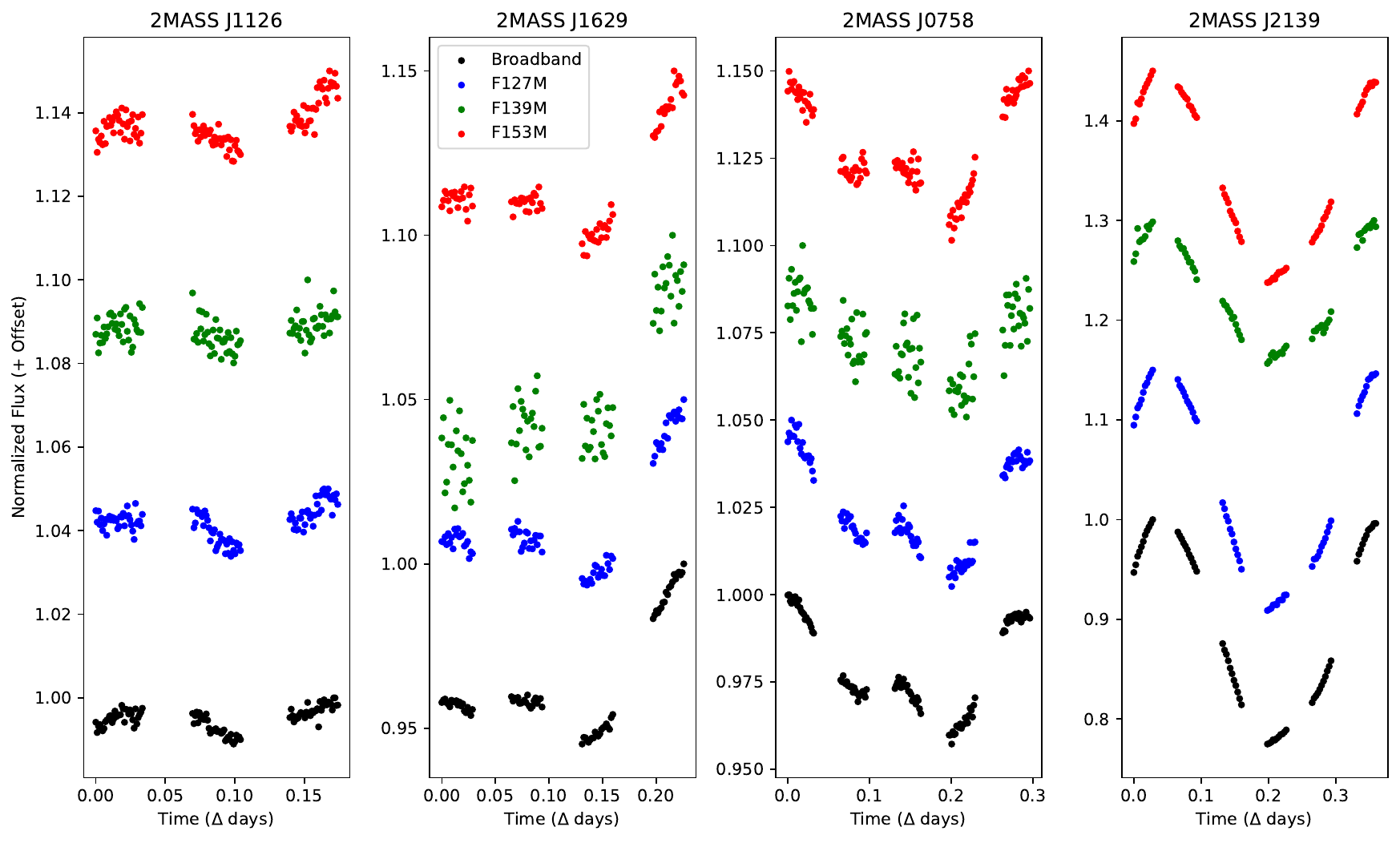} 
    \caption{Normalized light curves in four photometric bands: G141 broadband (1.10--1.65 \micron{}, black), F127M (blue), F139M (red), and F153M (green). Light curves are vertically offset for clarity: 0.05 flux units for the three leftmost targets (2MASS J1629, 2MASS J0758, and 2MASS J1126) and 0.15 flux units for 2MASS J2139 due to its larger variability amplitude.}
    \label{fig:lightcurves}
\end{figure*}

We applied synthetic photometry (\texttt{pysynphot}, \citealt{STScIDevelopmentTeam2013}) to generate light curves in four bands: the G141 broadband (1.10--1.65 $\mu$m), F127M ($J$ band peak, centered at 1.27 $\mu$m), F139M (water absorption band, centered at 1.39 $\mu$m), and F153M (centered at 1.53 $\mu$m) (Figure~\ref{fig:lightcurves}). The sinusoidal light curve of 2MASS J1126 yields a rotation period of $P=3.08 \pm 0.05$ hr. The period and uncertainty for 2MASS J1126 were determined using a least squares fit to a sinusoid model, and this measurement is consistent with $P=3.2$\,hr found by \cite{Metchev2015}. Additionally, \cite{Apai2013} detect $P=7.83$\,hr in their analysis of 2MASS J2139. In contrast, 2MASS J0758 and 2MASS J1629 exhibit irregular light curves that preclude period determination.

Colors (F127M-F139M, F139M-F153M, and F127M-F153M) as a function of time are also examined. With the exception of 2MASS J2139, no notable color change was detected. This matches our observations in Figure \ref{fig:minmaxratio} in which 2MASS J2139 is the only object that has clear wavelength-dependent variability. 

 \section{Atmospheric Modeling} \label{modeling}

\subsection{Model Methods and Assumptions}

    \begin{figure*}[t]
        \centering
        \includegraphics[width=\textwidth]{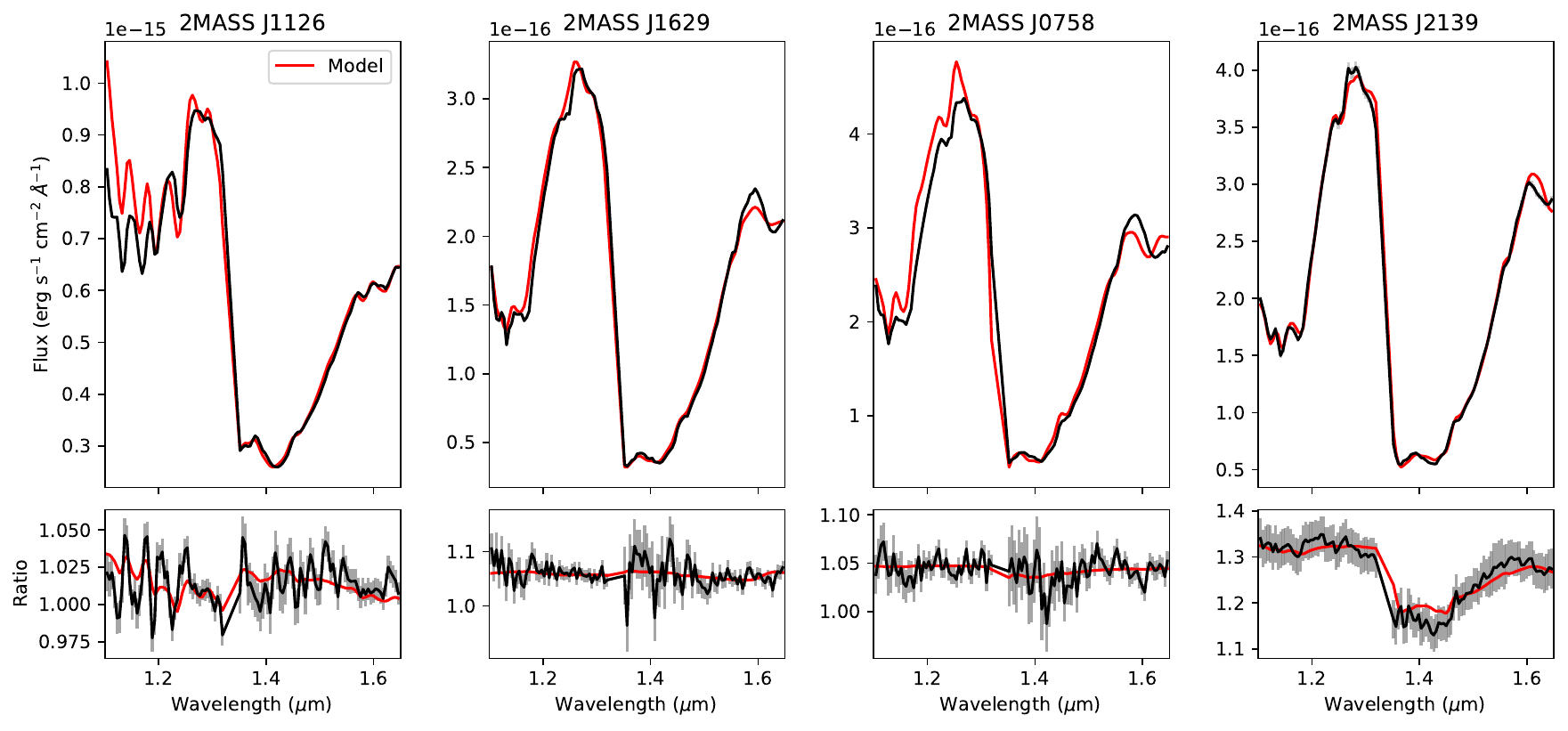} 
        \caption{\textit{Top panel:} The best-fit model spectrum $\fbase$ (red) and the average observed spectrum (black) of each object. The $\teff$ and $\fsed$ values that make up each model can be found in Table \ref{tab:modelfits}.
        \textit{Bottom panel:} The observed ratio and uncertainties (black) and the model ratio (red). The model ratio is calculated as \(\fbase/\ftot\) for all objects. We masked out the region where the flux has a steep slope ($1.32-1.35 \mu$m) for all objects.}
        \label{fig:modelspectra}
    \end{figure*}

We compare atmospheric models with observed spectra and variability to constrain surface temperature variations and cloud properties. We use the SONORA Diamondback grid because it represents the latest model development for incorporating condensate clouds \citep{Morley2024} \footnote{The thermal profiles are part of the model product provided by \citet{Morley2024}. \citet{Morley2024} noted the convergence challenge in several models' thermal profiles at low pressure levels. Those issues do not impact the wavelength ranges discussed here.}. These clouds are prevalent in L and L/T transition type objects in substellar atmospheres \citep[e.g.,][]{SuarezMetchev2022,SuarezMetchev2023}. The model grid sampled effective temperatures ($\teff$) from 900--2400 K (100 K steps), surface gravity ($\log g$) from 3.0 to 5.5 (0.5 dex steps), metallicity ($Z=$ 0, $\pm 0.5$), and sedimentation efficiency \fsed\ = 1, 2, 3, 4, 8, and cloud-free cases.

 


We assume the observed variability originates from the brown dwarfs' heterogeneous atmospheres with spatial variations in temperature and cloud coverage.
Following \cite{Marley2010} and \cite{Miles2023}, we construct the model atmosphere using a linear combination of two models, referred to as $\fbase$ and $\fpatch$: 
\begin{equation}\label{linearcomb}
     \ftot = (1-\alpha)\fbase+{\alpha}\fpatch,
\end{equation}
in which $\fbase$ and $\fpatch$ represent the larger and smaller area, respectively, and both are defined by specific $\fsed$ and $\teff$ values. $\alpha$ represents the covering fraction of $\fpatch$, ranging between 0 and 1. 
The flux ratio of the two extrema is modeled as $\ftot/\fbase$ (assuming bright spot, i.e. $\fpatch >\fbase$ ) or $\fbase/\ftot$ (assuming dark spot, i.e. $\fpatch<\fbase$). The $\fbase$ model is determined by fitting the time-averaged spectrum. The $\fpatch$ model and the covering fraction $\alpha$ are determined by the flux ratio between the maximum and minimum spectra.

\subsection{Fitting the Base Model}

To determine $\fbase$, we use the spectral fitting package \texttt{species} \citep{Stolker2020} to find the best-fitting scaling factor, \teff, $\log g$, metallicity, and $f_\mathrm{sed}$. To begin the fitting process, we linearly interpolated the models onto the HST wavelength grid to match its spectral resolution. The model is convolved with a instrument profile corresponding to the spectral resolution of WFC3/G141 ($R=130$) before comparing to the data. We introduce an error inflation parameter that uniformly scales all uncertainties. This accounts for both underestimated observational errors and systematic uncertainties in the model. The parameter space is explored using a nested sampling method with 500 live points to maximize the likelihood function \citep[][]{Feroz2009,Buchner2014}. For all four cases, the fittings converge to best-fitting solutions. Table~\ref{tab:modelfits} lists the best-fitting parameters. In the appendix, Figure~\ref{fig:cornerplots} illustrates the distributions of the parameters.


Spitzer IRS spectra directly probe silicate absorption near 10\,\micron{}. We experimented with including archival Spitzer IRS data ($\sim5.2-14.1\,\mu$m) on 2MASS J2139, J1126, and J0758 provided in \citet[][]{SuarezMetchev2022,SuarezMetchev2023} when fitting for $\fbase$. In all cases, inclusion of the Spitzer spectra does not significantly change the best-fitting values listed in Table~\ref{tab:backgroundinfo}. We do not include the Spitzer data for the rest of the paper for two reasons: first, the clouds that impact the HST observations are at altitudes different from those introducing the silicate features in the Spitzer spectra \citep{Luna2021}.  Second, the HST and Spitzer observations were conducted at different epochs, and thus the atmospheric structures determined by HST are likely different from those found in the Spitzer data.

All four fits converge to high metallicity solutions. However, our data cannot meaningfully constrain metallicity given the low spectral resolution and narrow wavelength coverage. We interpret this as a numerical artifact and do not discuss it further.



\begin{deluxetable}{l|cccc|ccc|c}
\tabletypesize{\footnotesize}
\tablecaption{Best-fit Heterogeneous Atmospheric Model Parameters \label{tab:modelfits}}
\tablehead{
\colhead{Brown Dwarf} & \multicolumn{4}{c}{Base Flux} & \multicolumn{3}{c}{Patch Flux} & \colhead{$\alpha$}\\
\colhead{} & \colhead{$T_{\mathrm{eff}}$ (K)} & \colhead{$f_{\mathrm{sed}}$} & \colhead{$\chi^2_0$} &  \colhead{$\ln(z)$} & 
\colhead{$T_{\mathrm{eff}}$ (K)} & \colhead{$f_{\mathrm{sed}}$} & \colhead{$\chi^2_0$} & \colhead{} 
}
\startdata
2MASS J1126  & 1562 & 7.53 & 8320 & 3955 & 1500 & 8 & 1.96 & 0.063\\
2MASS J1629  & 1243 & 7.84 & 390 & 4140 & 900 & 4 & 0.80 & 0.067\\
2MASS J0758  & 1230 & 7.43 & 3530 & 4005 & 900 & 2 & 0.83 & 0.047\\
2MASS J2139  & 1227 & 5.50 & 19 & 4109 & 1000 & 1 & 0.34 & 0.263\\
\enddata
\end{deluxetable}


\subsection{Fitting the Max/Min Flux Ratio}

We used least-$\chi^2$ fitting techniques to determine both the fractional size ($\alpha$) and the best-fitting $\fpatch$ model. We did not use \texttt{species} or other similar packages for this step because heterogeneous atmospheric models have not been supported. Unlike in the calculation of $\fbase$, $\alpha$ is a free parameter in our calculation of $\fpatch$, and there are 116 total degrees of freedom for each object. We determine $\fpatch$ and $\alpha$ by minimizing the $\chi^2$ value in the equation below. We also force the $\fbase$ and $\fpatch$ models to share the same surface gravity and metallicity.

 \begin{equation}\label{eq:avg_chi2}
     \chi^2=\sum_i\frac{(O_i-M_i)^2}{\sigma_i^2},
 \end{equation}

 in which $O_i$ represents the observed flux ratio, $M_i$ represents the model flux ratio ($\fbase/\ftot$ or $\ftot/\fbase$). The uncertainty term ($\sigma$) in Equation \ref{eq:avg_chi2} represents the propagated uncertainty of the observed flux ratio, with individual flux uncertainties combined in quadrature.
 


\begin{figure*}
    \centering

    \parbox[c]{5cm}{Optimizing $\fpatch$ Parameters} 
    {\includegraphics[width=\textwidth]{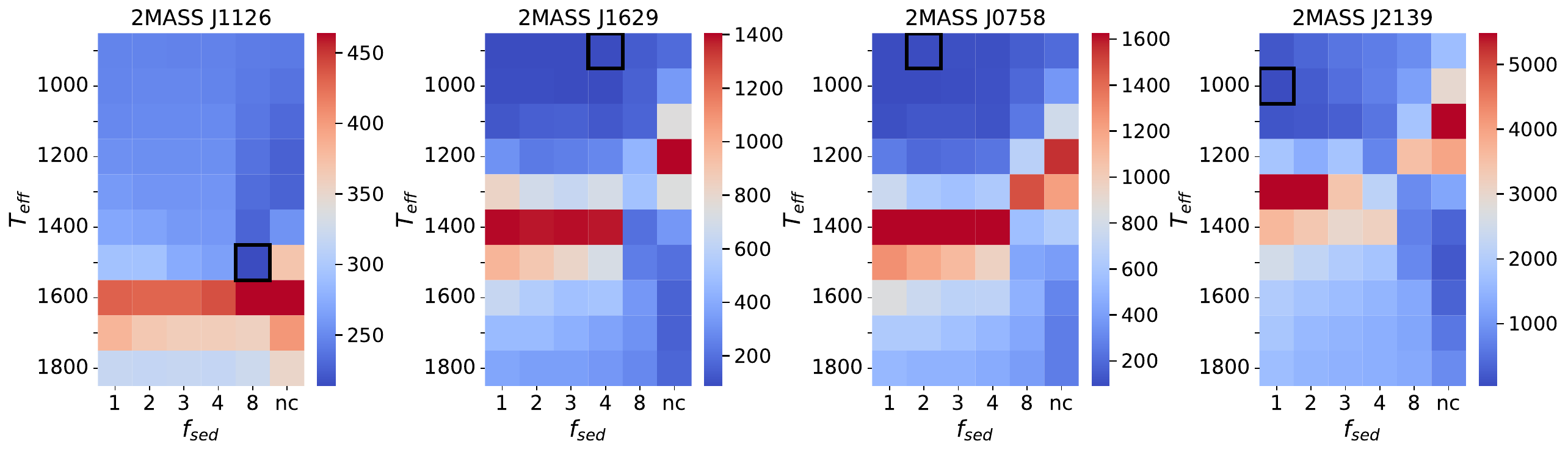}}

    \caption{The minimum $\chi^2$ values as functions of the $\fsed$ vs. $\teff$ for $\fpatch$. The overall minimum $\chi^2$ values for the map is highlighted using a black rectangle, and the highest and lowest $\chi^2$ values are represented by the reddest and bluest colors, respectively. 
    }
    \label{fig:chi_sq_min}
\end{figure*}

The heat maps in Figure \ref{fig:chi_sq_min} depict the minimum $\chi^2$ values for each respective combination of $\fsed$ vs. $\teff$ for $\fpatch$. The overall best-fit $\teff$ and $\fsed$ values and $\alpha$ fraction that make up each $\fpatch$ model can be found in Table \ref{tab:modelfits}, along with their corresponding reduced $\chi^2$ values.
Additionally, the average observed spectrum and the best-fit model spectrum $\fbase$ are plotted alongside the observed and model flux ratios in Figure \ref{fig:modelspectra}.


The observed spectral variability for the three early-T type objects are best explained by heterogeneous cloudy models. The base and patch models are both cloudy and differ in their effective temperature.
Of the $\ftot$ model combinations, 2MASS J1629 and 2MASS J0758 have strikingly similar atmospheric structures. Their spectral variability is best explained by small patches with cooler $\teff$ and thicker clouds. The large $\alpha$ fraction of 2MASS J2139 relative to the other targets agrees with its large variability amplitude.
For 2MASS J1126, the best-fitting base and patch model have the same $\fsed$ values and differ in $\teff$. This suggests its spectral variability is driven by mechanisms different from those seen in the three early-T objects.
Additionally, our model results for 2MASS J1126 are consistent with those in \cite{Burgasser2008}, which probed both the optical and near-infrared and estimated an atmosphere of $\teff$ = 1700K and $\fsed$ = 4 by comparing the spectrum of 2MASS J1126 to other optically classified mid L dwarfs. They concluded that thin condensate clouds were the cause of 2MASS J1126’s bluer near-infrared spectrum relative to similar spectral types. We find a similar base temperature and cloud thickness ($\teff\sim1500$K  and $\fsed=8$). 

\section{Color Magnitude Evolution Analysis} \label{cmd}

To determine the dominant mechanism that drives spectral variability in each brown dwarf, we compare the observations of variable brown dwarfs with both empirical data and models on color-magnitude diagrams.

\subsection{Comparing to Models on Color Magnitude Spaces}

\begin{figure}
    \centering
    \includegraphics[width=1\linewidth]{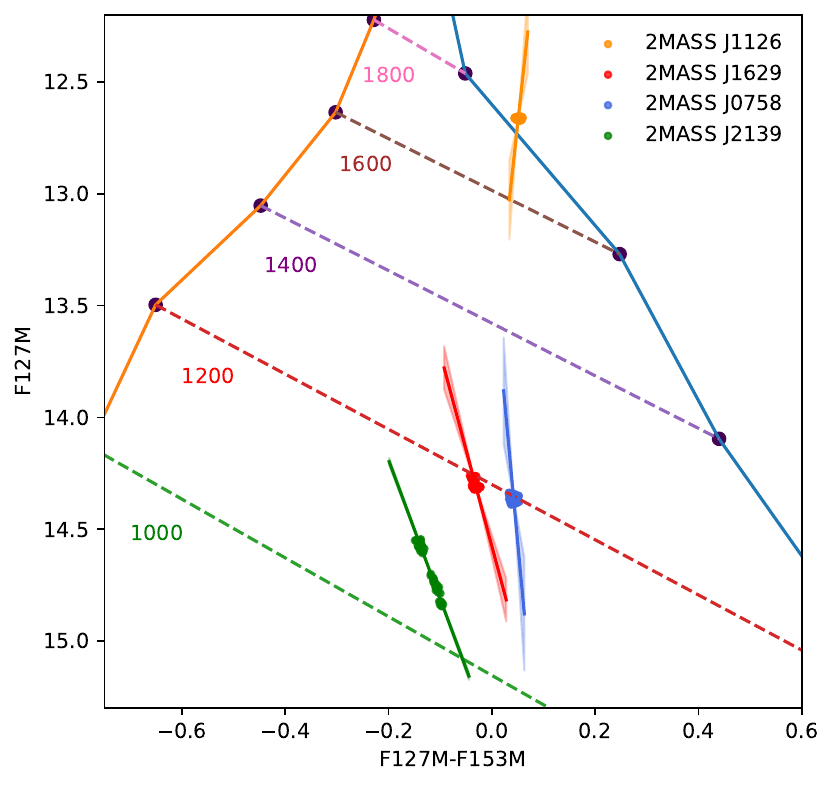}
    \caption{Color magnitude diagram for F127M-F153M vs. F127M and the evolution of each target plotted with their respective lines of best fit and model tracks derived from SONORA Diamondback spectral grid. The labeled dashed lines represent constant $\teff$, and the blue solid line in both panels corresponds to no sedimentation parameter (cloud-free) and the orange solid line corresponds to $f_{sed}$=1 (thick clouds). The best-fitting lines only indicate the directions of the color-magnitude variations, not the amplitudes. 2MASS J1126, J1629, J0758, and J2139 are shown in orange, red, blue, and green, respectively.}
    \label{fig:f127mf153m}
\end{figure}

\begin{figure*}
    \centering
    \includegraphics[width=\linewidth]{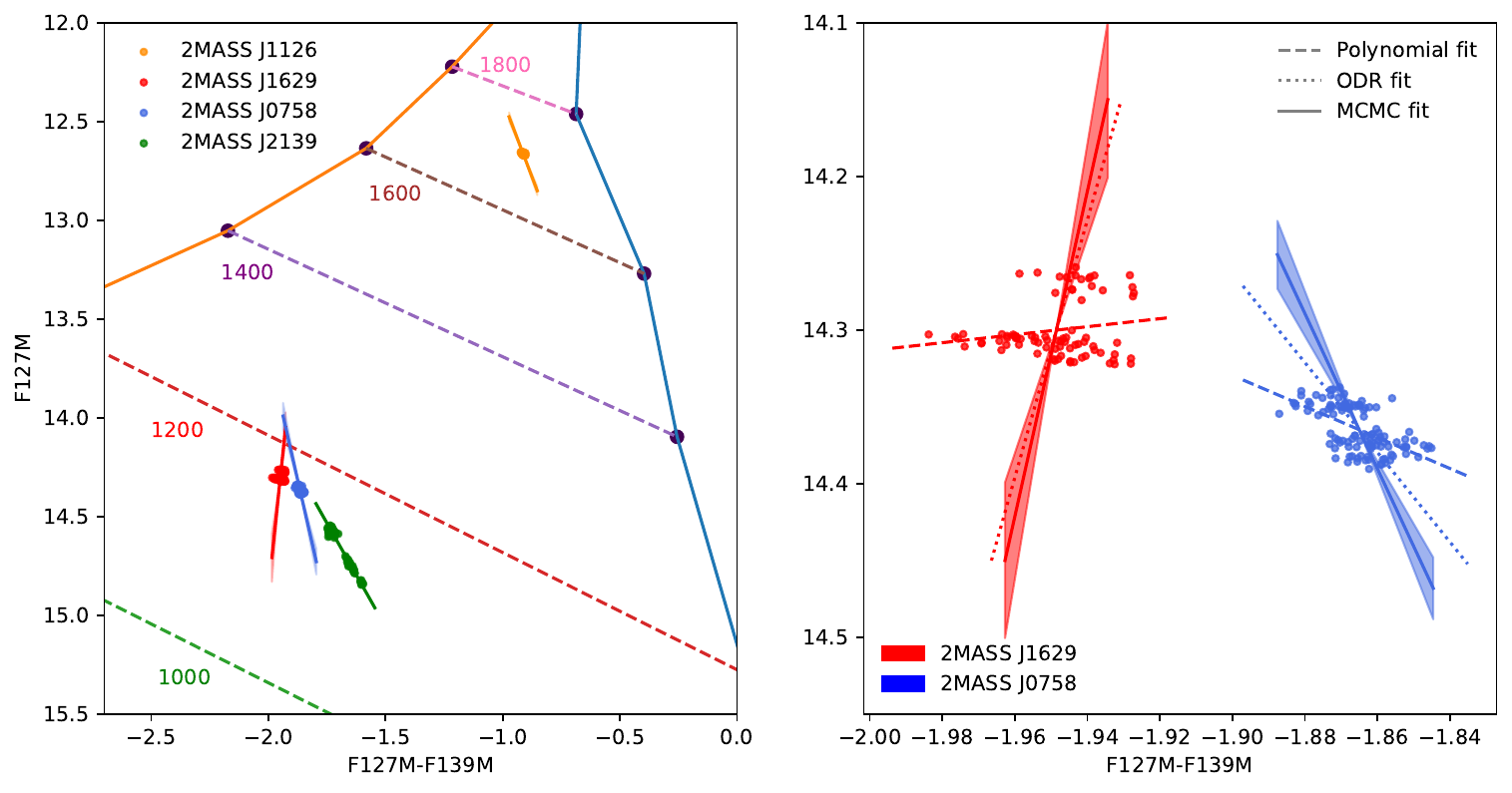}
    \caption{\textbf{Left:} F127M$-$F139M vs. F139M color-magnitude diagram. Best fitting lines and their uncertainties (shaded regions) to the observe color-magnitude variations
     are shown for all objects, alongside SONORA Diamondback model tracks for various temperatures (dashed colored lines). Cloud models include cloud-free (blue solid) and thick clouds with $f_{\text{sed}}=1$ (orange solid). Models with other $\fsed$ values fall between the orange and blue lines and are not shown. The lengths of the best-fitting lines only indicate the directions of the color-magnitude change, not the strengths. 2MASS J1126, J1629, J0758, and J2139 are shown in orange, red, blue, and green, respectively. \textbf{Right:} A comparison of fitting methods for 2MASS J1629 (red) and 2MASS J0758 (blue) on the F127M$-$F139M vs. F139M diagram. 
     We adopt MCMC as our final fitting method because it accounts for robust and empirically estimated errors in both axes, and we overlay the estimated slope uncertainties (shaded regions).}


    \label{fig:f139m}
\end{figure*}

We transform the model spectral grid into color-magnitude space by deriving their synthetic photometric data using the pysynphot package \citep{STScIDevelopmentTeam2013}: the same method as used for the individual spectra of our targets. We then created two color-magnitude diagrams -- F127M-F153M vs. F127M and F127M - F139M vs. F127M (Figures \ref{fig:f127mf153m} and \ref{fig:f139m}). An object's slope in these diagrams reflects the mechanisms driving its variability. Variability driven by patchy clouds would correspond to shifts more aligned with the $\fsed$ grid lines, while variability driven by hot spots would move objects along the temperature grids.
Qualitatively comparing the trajectory of the observed targets with the directions of the cloud thickness and temperature variations predicted by the model grid can inform us of the mechanisms driving each object's variability.






Accurate best-fit slopes are essential for determining variability mechanisms, particularly for 2MASS J1629 and 2MASS J0758, whose small variability makes trajectories highly method-dependent. The right plot in Figure \ref{fig:f139m} compares three fitting approaches: the \texttt{numpy} \texttt{Polynomial.fit} method (dashed lines); the \texttt{scipy.odr} orthogonal distance regression (dotted lines); 
and the MCMC sampling (implemented with \texttt{emcee} \citet{Foreman-Mackey2013}) of a linear model (solid lines). We find the last approach, emcee fitting, to be the most robust one and adopt it for all targets across all color-magnitude diagrams, as it accounts for empirically estimated errors in both axes. To estimate the uncertainty in absolute magnitude from the MCMC sampling, we applied a median filter to remove astronomical noise, subtracted these results from our observations, and applied the standard deviation of the output to each data point to represent the uncertainty. From there, we used basic error propagation to find the uncertainty in color for each point. Table \ref{tab:cmdslopes} provides the best-fitting slopes and uncertainties using the MCMC method.

\begin{figure}
    \centering
    \includegraphics[width=1\linewidth]{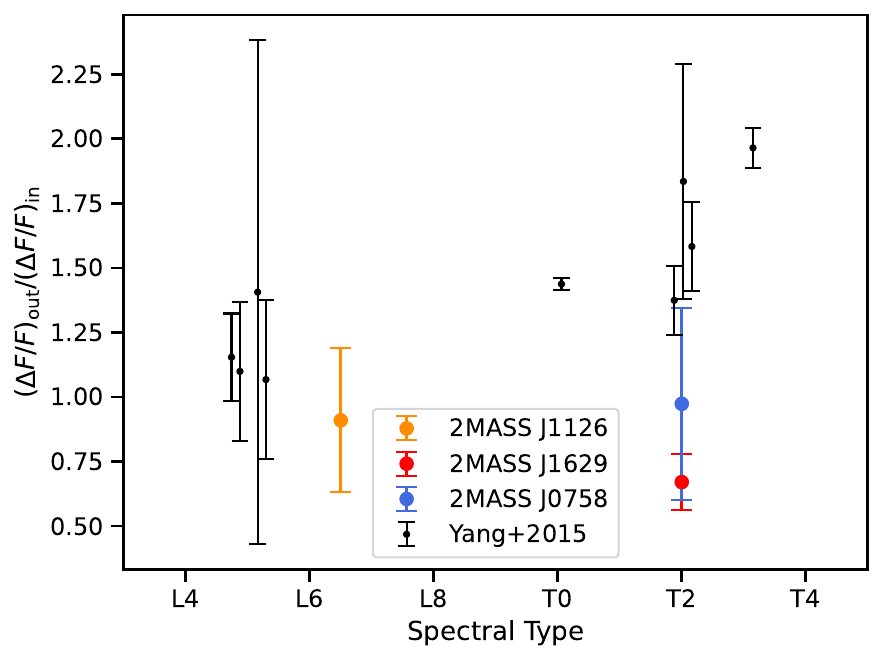}
    \caption{A comparison of in- and out-of-water absorption variability amplitude between brown dwarfs presented in this work and those published by \citet{Yang2015}. 2MASS J0758 and 2MASS J1629 stand out because they do not show muted variability in the water absorption bands, which are prevalent in early-T type brown dwarfs.}
    \label{fig:yangreplication}
\end{figure}

\begin{deluxetable}{ccc}  
\tablecaption{Observed Slopes for both CMDs \label{tab:cmdslopes}}  
\tablehead{  
\colhead{Object} & 
\colhead{\shortstack[t]{\rule{0pt}{2.5ex}F127M--F153M \\ vs. F127M slope}} &
\colhead{\shortstack[t]{\rule{0pt}{2.5ex}F127M--F139M \\ vs. F127M slope}}
}  
\startdata  
2MASS J1126 & -21.1 $\pm$ 10.3 & 3.1 $\pm$ 0.4 \\  
2MASS J1629 & 8.6 $\pm$ 1.6 & -11.3 $\pm$ 3.4 \\  
2MASS J0758 & 24.9 $\pm$ 12.3 & 5.2 $\pm$ 0.9 \\  
2MASS J2139 & 6.2 $\pm$ 0.2 & 2.1 $\pm$ 0.0 \\  
\enddata  
\end{deluxetable}

\subsubsection{Variability In- and Out- of the Water Absorption Band}

The trajectories of the three early T dwarfs are broadly consistent across both color-magnitude diagrams. Based on the best-fit slope relative to the model grid in Figure \ref{fig:f127mf153m}, changing both cloud thickness and effective temperature is required to explain the color-magnitude trajectories of these brown dwarfs.
However, the majority of 2MASS J2139, J0758, and J1629's observed data points lie between the 1000K and 1200K model temperature lines, which are slightly lower than the $\teff$ of the best-fitting base spectrum in Table \ref{tab:modelfits}. 

In the F127M vs.\ F127M$-$F139M CMD (Figure \ref{fig:f139m}), which combines continuum and water absorption filters to enhance sensitivity to variability mechanisms \citep{Morley2014}, the three T dwarfs exhibit different behaviors. 2MASS J2139 shows high-amplitude variability primarily driven by cloud thickness variation, as evidenced by the alignment between its CMD trajectory and the cloud-varying model track. In contrast, 2MASS J1629's CMD track demonstrates a mostly gray color change, while 2MASS J0758's slope falls between those of 2MASS J2139 and J1629. Unlike well-studied high-amplitude variable early-T brown dwarfs such as 2MASS J2139 and SIMP0136, whose color changes in the 1.4\,$\mu$m water absorption band are driven by forsterite clouds at the one bar pressure level \citep{Vos2023}, the water-band color changes in 2MASS J1629 and 2MASS J0758 cannot be attributed to a single dominant mechanism. Moreover, the steep slope of 2MASS J1126 in both color-magnitude diagrams is consistent with its gray variability in Figure \ref{fig:minmaxratio}.

Figure~\ref{fig:yangreplication} compares in- and out-of-water band variability amplitude ratio ($(\Delta F/F)_{out}/(\Delta F/F)_{in}$) for brown dwarfs targeted in this study with those presented by \cite{Yang2015}. They found an increasing trend of this amplitude ratio from mid-L to early-T dwarfs (Figure \ref{fig:yangreplication}), noting that early-T dwarfs exhibit a greater relative flux change inside the continuum. In contrast, our L/T transition objects 2MASS J0758 and 2MASS J1629 have a relatively constant relative flux change and deviate from the previously identified trend. Instead, high-altitude haze, previously identified as a source of gray variability in mid-L dwarfs \citep{Yang2015}, may extend to the early T-type and represent a promising explanation for the color changes observed in 2MASS J1629 and 2MASS J0758. The L6.5 dwarf 2MASS J1126 in our sample is consistent with this trend, as demonstrated by its steep slope in Figure \ref{fig:f139m}. These diverse variability properties highlight the need for sample studies using broad wavelength coverage to monitor multiple molecular features and atmospheric pressure levels in L/T transition brown dwarfs, which extends the approach of recent case studies \citep[e.g.,][]{Biller2024,McCarthy2025,Chen2025}.

\subsection{Comparing to an Empirical Color-Magnitude Diagram}

\begin{figure}
    \centering
    \includegraphics[width=1\linewidth]{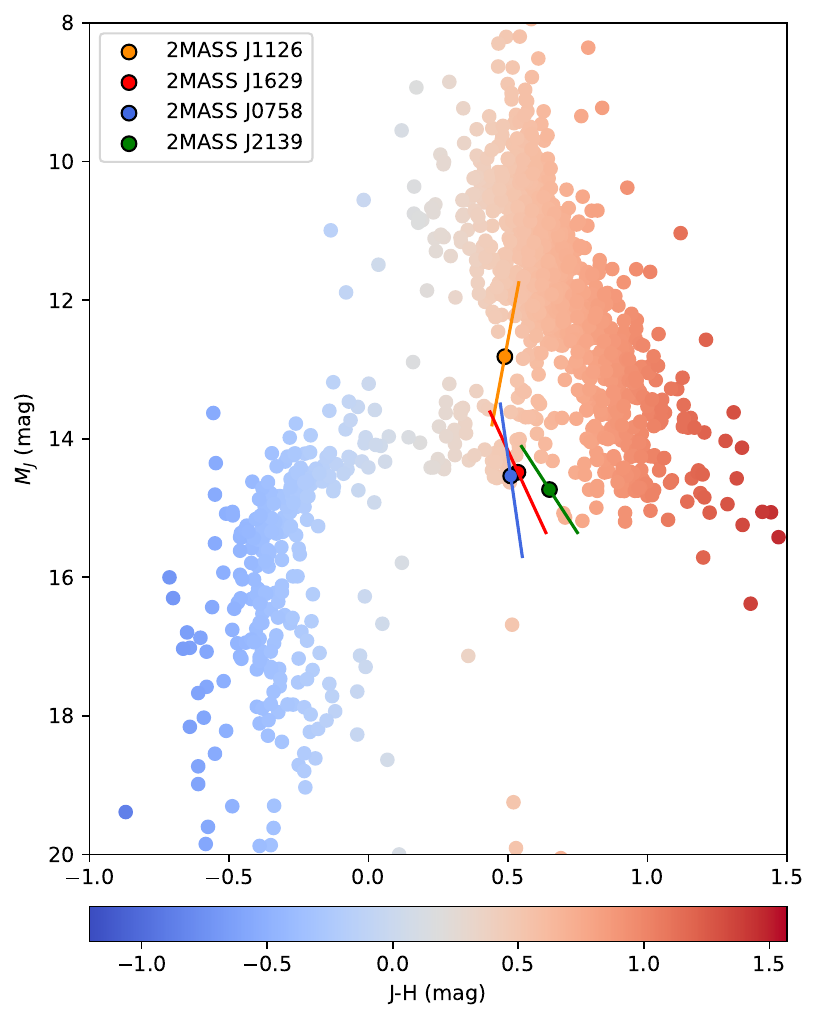}
    \caption{Color magnitude diagram MKO $J-H$ vs. $M_J$ of our four targets compared to the other known late M- to early Y- dwarfs using data from the Ultracool sheet \citep{Best2024-oc}. 
    The lines plotted with each object represent the best-fit slope to their F127M-F153M vs. F127M variations (Figure \ref{fig:f127mf153m}). 2MASS J1126, J1629, J0758, and J2139 are shown in orange, red, blue, and green, respectively.}
    \label{fig:ultracool_cmd}
\end{figure}


Figure \ref{fig:ultracool_cmd} is a MKO J-H color vs. $M_J$ magnitude diagram that plots empirical data of 3890 known brown dwarfs, whose photometry and distance values are obtained from the Ultracool Sheet \citep{Best2024-oc}. Our targets are highlighted and labeled relative to the other L- and T- dwarfs. 
Because our wavelength range does not span the entirety of the H-band, we are not able to plot our observations on the empirical CMD. However, we tracked target evolution across frames by plotting each in F127M-F153M color versus F127M magnitude space and calculating best-fit slopes. These slopes, centered on each target, were overlaid on our empirical color-magnitude diagram. This approach approximates the color changes of $J-H$ with those of F127M$-$F153M. We note that G141's long wavelength cutoff prevents complete H-band flux recovery.

The three early T dwarfs we study appear to have a similar color-magnitude evolution in our empirically derived color-magnitude diagram. This conclusion is consistent with that of \cite{Lew2020}, which similarly detects 12 L/T transition objects evolving more on a vertical scale in color J-H vs. magnitude M$_J$ space. Our results reinforce the finding that L/T transition color evolution and individual object variability arise from distinct underlying physical mechanisms.

\section{Summary} \label{summary}
We present new HST/WFC3 spectral time-series observations of three variable brown dwarfs. These observations offer insights into their variability, atmospheric composition, and cloud coverage. The findings of our investigation are as follows.

\begin{itemize}
    \item Our observations yielded high-precision spectra with signal-to-noise ratios ranging from 50 to 150 per frame per unit wavelength for each target. 
    \item We detected significant spectroscopic variability in the three new targets in the 1.1 to 1.65 micron wavelength range and recovered the published variability of 2MASS J2139. The measured variability amplitudes are 1.12\%, 5.80\%, and 4.47\% for 2MASS J1126, 2MASS J1629, and 2MASS J0758, respectively. The wavelength dependence on variability of these three objects is not as significant as previous observations of 2MASS J2139. 
    \item Broadband, F127M, F139M, and F153M light curves were calculated to evaluate periodicity and study the color dependence of the variability. For 2MASS J1126, we determined a period of 3.09 hr. Due to their irregular light curves, we cannot detect a precise period measurement for 2MASS J0758 and 2MASS J1629. 
    \item We construct heterogeneous atmospheric model by linearly combining SONORA Diamondback spectra that vary in effective temperature and cloud opacity  \citep{Morley2024} and fit the average observed spectrum and observed ratio using the least $\chi^2$ technique. 
    Our heterogeneous atmospheric model reproduces the observed spectra and wavelength-dependent variability.
    \item All three early T dwarfs exhibit broadly gray variability in the F127M-F139M vs. F127M color-magnitude space, but with distinct trajectories that reflect differences in their atmospheric properties. For example, 2MASS J2139’s trajectory points to patchy clouds as the dominant driver, 2MASS J1629’s gray color change suggests contributions from both patchy clouds and temperature variations, and 2MASS J0758 falls between these cases. This result encourages future observations that allow the comparisons of multiple objects across various color-magnitude diagrams.
    \item As a mid-L type comparison source, 2MASS J1126 also demonstrates gray variability patterns, consistent with those found in brown dwarfs with similar spectral types. The similarity in variability color-dependence between 2MASS J0758, J1126, and J1629 suggests that mechanisms explaining gray variability in mid-L types may extend to early-T types.

\end{itemize}

\begin{acknowledgments}
We thank the referee for a prompt and constructive report, which has helped significantly improve the quality and rigor of the manuscript. We acknowledge support from the Virginia Initiative for Cosmic Origins (VICO) summer program, during which portions of this work were completed. This research is based on observations made with the NASA/ESA Hubble Space Telescope obtained from the Space Telescope Science Institute, which is operated by the Association of Universities for Research in Astronomy, Inc., under NASA contract NAS5–26555. Y.Z. acknowledges HST data analysis grants associated with programs GO-16036. M.C. and Y.Z. acknowledge support from Heising-Simons Foundation 51 Pegasi b Alumni Faculty Grant (2023-4808 – 51).
\end{acknowledgments}

\software{
Astropy \citep{astropy:2013, astropy:2018, astropy:2022},
NumPy \citep{harris2020array},
SciPy \citep{virtanen2020scipy},
matplotlib \citep{hunter2007matplotlib},
Seaborn \citep{waskom2021seaborn},
emcee \citep{foremanmackey2013emcee},
corner.py \citep{corner},
PySynphot \citep{STScIDevelopmentTeam2013},
species \citep{Stolker2020},
pyMultinest \citep{Buchner2014},
ultranest \citep{Buchner2021}
}

\bibliographystyle{aasjournal}
\bibliography{sample631}

\appendix

\section{Appendix A: Corner Plots}

\begin{figure}[htbp]
    \centering
    \begin{subfigure}[b]{0.49\textwidth}
        \centering
        \includegraphics[width=\textwidth]{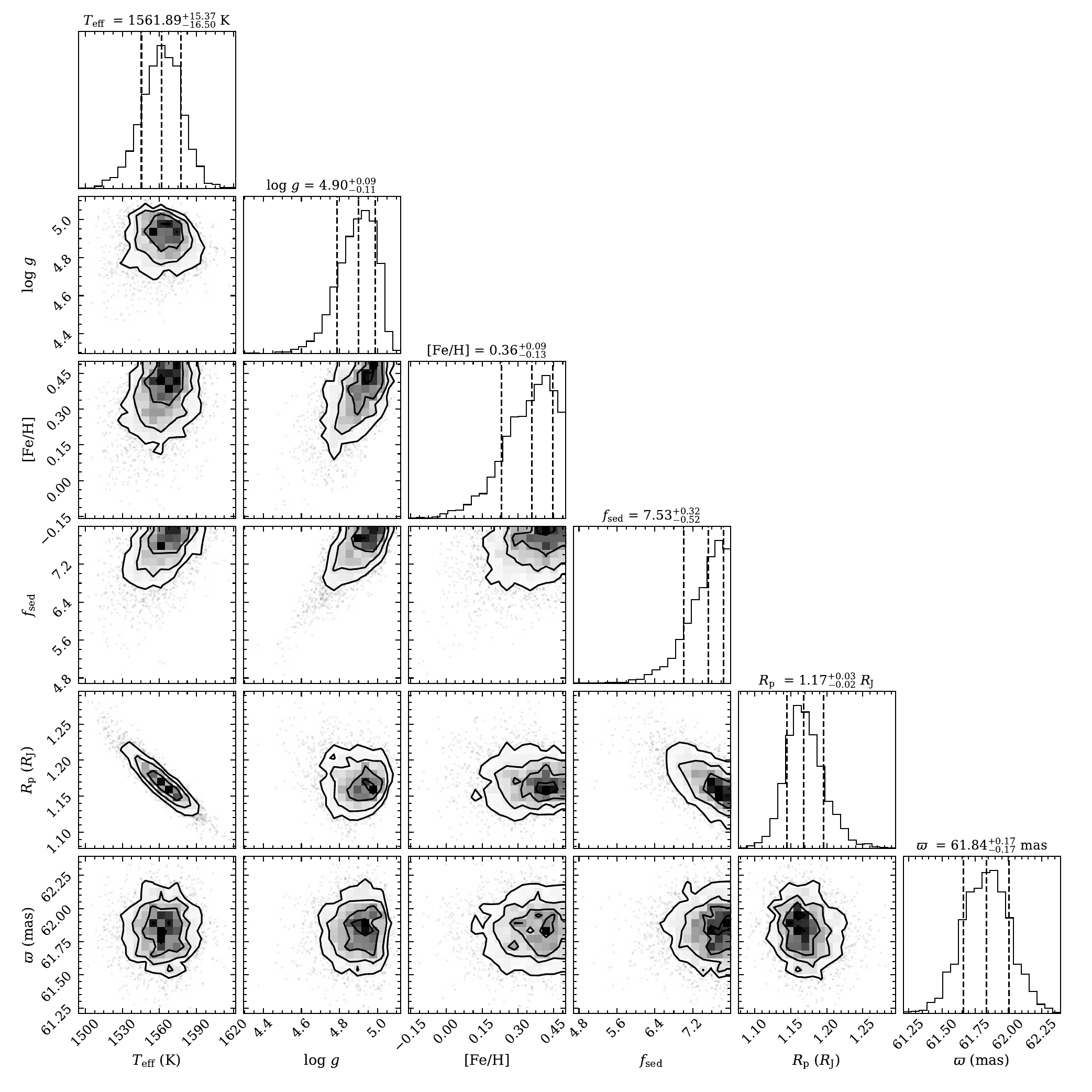}
        \caption{2MASS J1126}
        \label{fig:1126posterior}
    \end{subfigure}
    \hfill
    \begin{subfigure}[b]{0.49\textwidth}
        \centering
        \includegraphics[width=\textwidth]{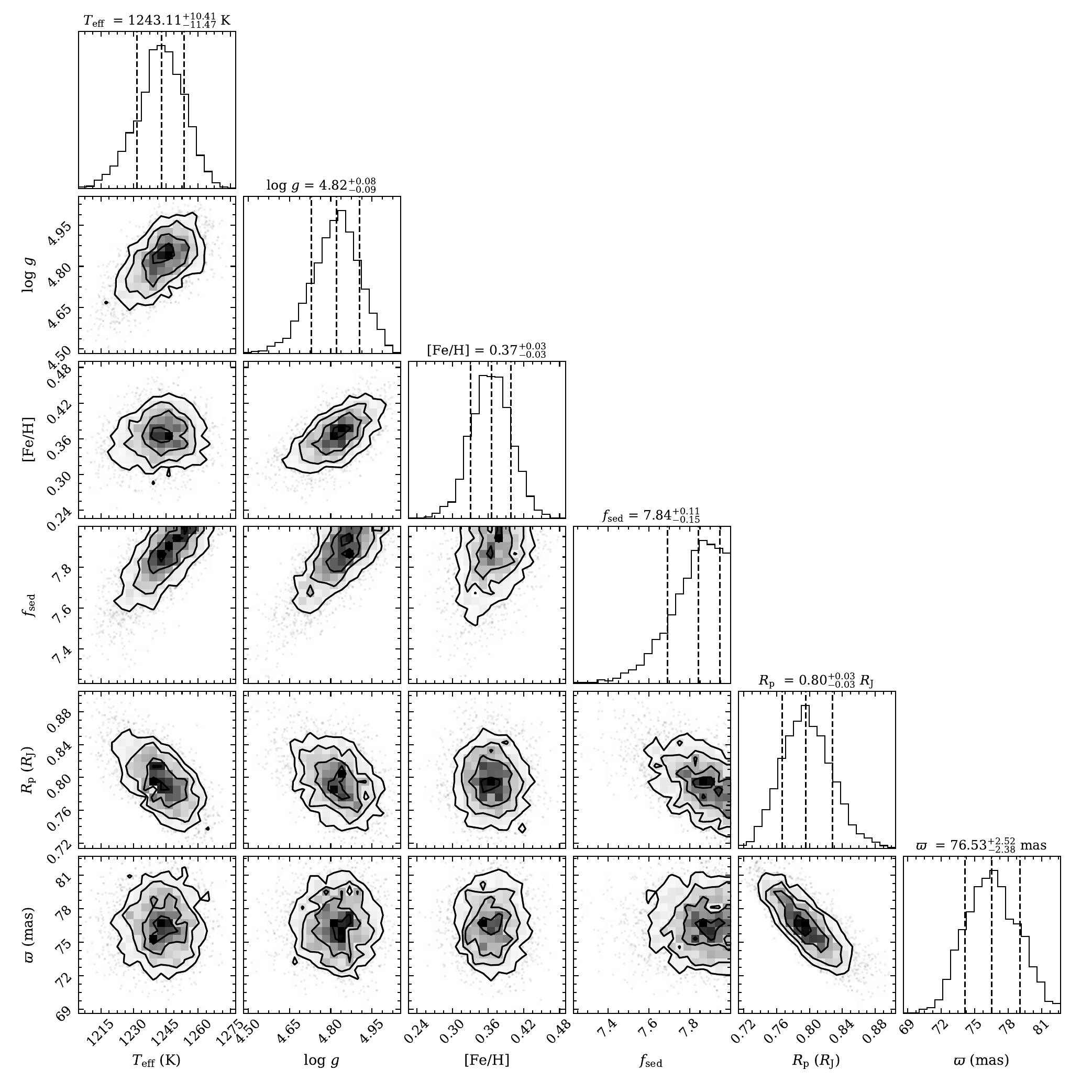}
        \caption{2MASS J1629}
        \label{fig:1629posterior}
    \end{subfigure}

    \vskip\baselineskip
    \begin{subfigure}[b]{0.49\textwidth}
        \centering
        \includegraphics[width=\textwidth]{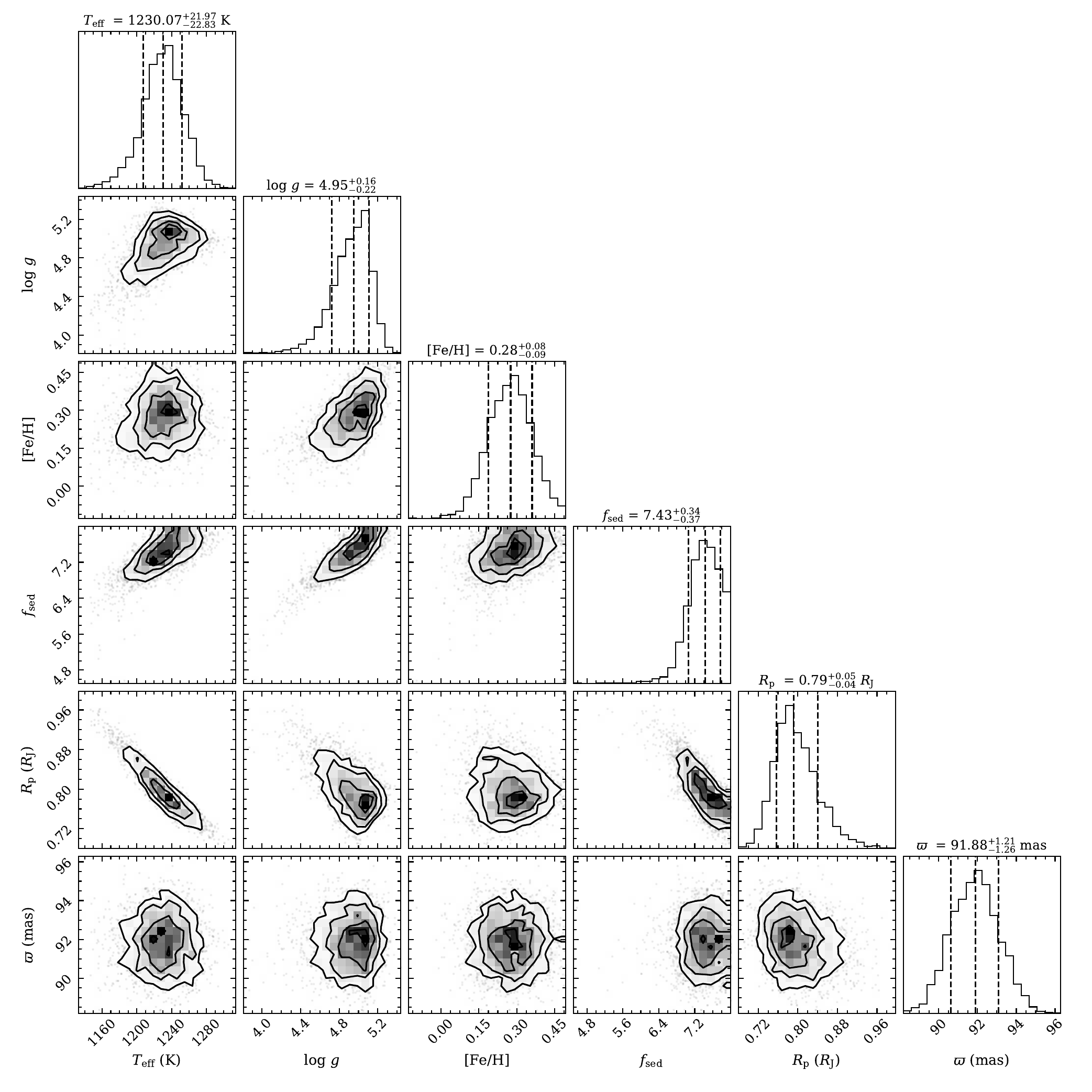}
        \caption{2MASS J0758}
        \label{fig:0758posterior}
    \end{subfigure}
    \hfill
    \begin{subfigure}[b]{0.49\textwidth}
        \centering
        \includegraphics[width=\textwidth]{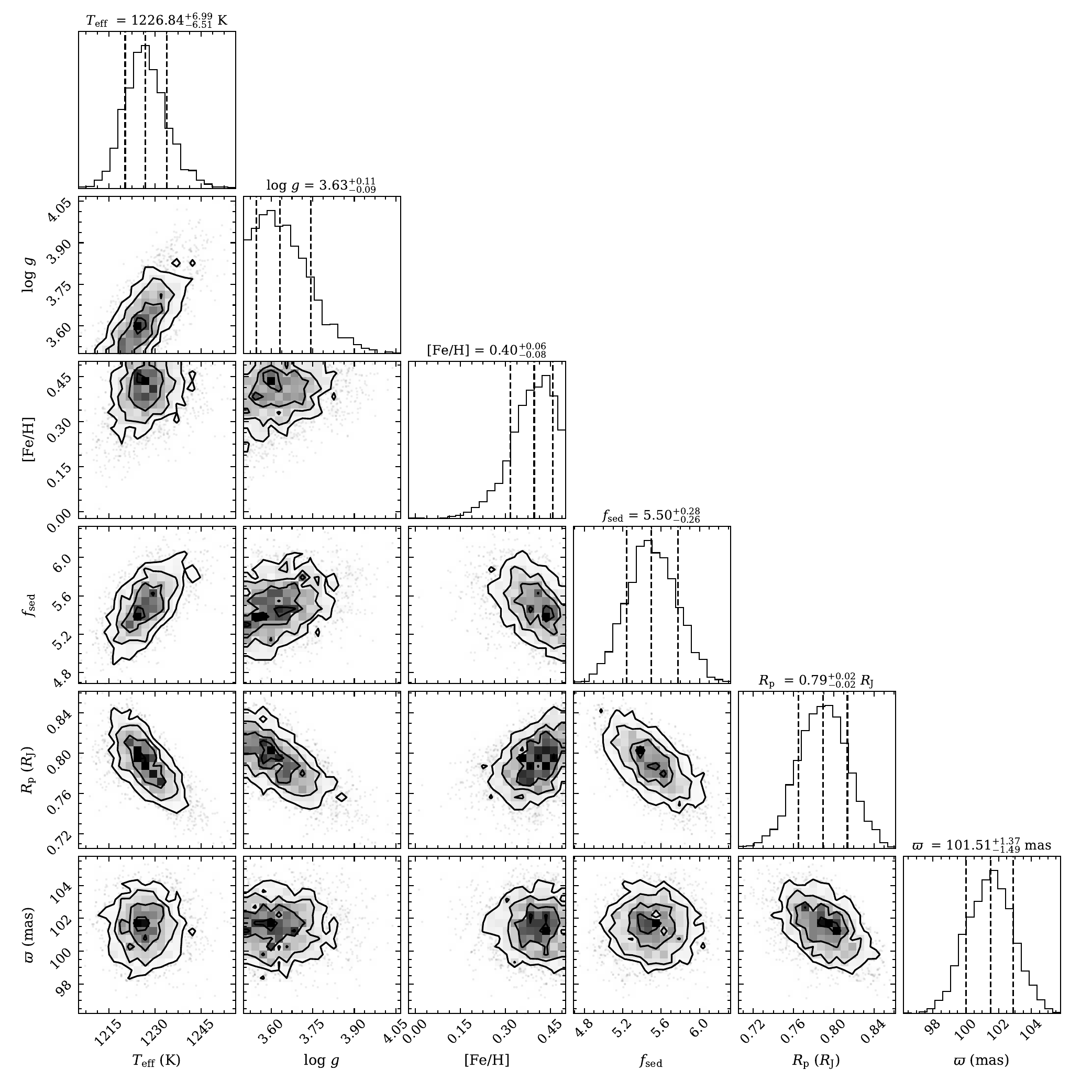}
        \caption{2MASS J2139}
        \label{fig:2139posterior}
    \end{subfigure}

    \caption{Corner plots from the $\fbase$ fits for all objects}
    \label{fig:cornerplots}
\end{figure}

\end{document}